Invited Article

# Figuring out Art History

Maximilian Schich

**Abstract:** World population and the number of cultural artifacts are growing exponentially or faster, while cultural interaction approaches the fidelity of a global nervous system. Every day hundreds of millions of images are loaded into social networks by users all over the world. As this myriad of new artifacts veils the view into the past, like city lights covering the night sky, it is easy to forget that there is more than one Starry Night, the painting by Van Gogh.

Like in ecology, where saving rare species may help us in treating disease, art and architectural history can reveal insights into the past, which may hold keys to our own future. With humanism under threat, facing the challenge of understanding the structure and dynamics of art and culture, both qualitatively and quantitatively, is more crucial now than it ever was. The purpose of this article is to provide perspective in the aim of figuring out the process of art history – not art history as a discipline, but the actual history of all made things, in the spirit of George Kubler and Marcel Duchamp. In other words, this article deals with the grand challenge of developing a systematic science of art and culture, no matter what, and no matter how.

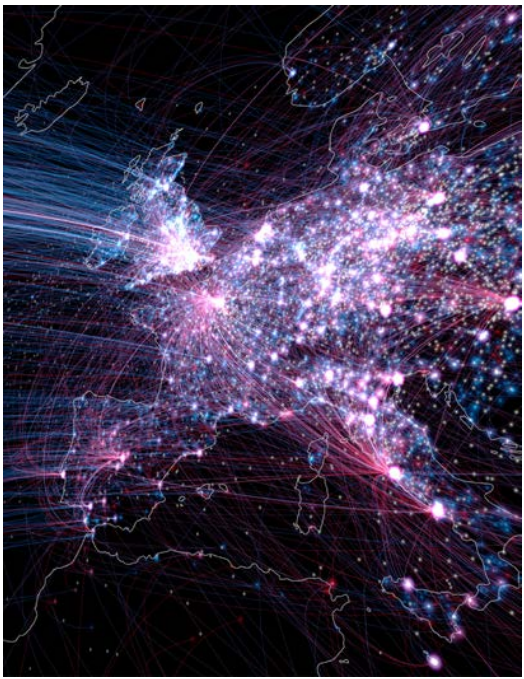

**Figure 1: Between this first visualization and publication lie three years of doing science.** The figure shows a still from an animation tracking individuals from birth to death (from blue to red). Published within "A Network Framework of Cultural History" in Science Magazine and in a poetic transformation as "Charting Culture" in the Nature video channel, the original purpose of the visualization was to help the group of researchers to find and understand quantitative patterns. The final animation eventually accumulated more than one million views, and was featured among "Best Data Visualizations in 2014" in FlowingData, "Best American Infographics 2015", the "NSF/PopSci Vizzies top 10", and "Macroscopes to Interact with Science" at Scimaps.org. Visualization: Maximilian Schich & Mauro Martino, www.cultsci.net.

# Introduction[1]

Imagine we knew there was life on Mars and there were biologists that are both equipped with the right skills and fearless enough to figure it out. It would be obvious to fund more research into Martian biology, including to build and send a spaceship to observe Martian life up close, at least from orbit. In art history, we are in a similar situation. We know there is cultural complexity emerging from local activity of large numbers of cultural agents. Based on qualitative observation and quantitative measurement we are familiar with intricate large-scale patterns in art and culture that are non-average, non-random, and hard-to-understand. While many traditional departments of art history are stagnant or shrinking, there are rapidly growing numbers of researchers in computer science and physics both curious and well equipped to advance our understanding of cultural complexity. Combining their curiosity and skills with solid domain expertise in the arts and culture, we are ready to build laboratories that will advance our understanding beyond the anecdotal, theoretic, or what can be achieved by even the most productive researchers using qualitative inquiry alone.

This article provides a perspective towards a systematic science of art and culture, where possible advances are driven by explosively growing amounts of data, including images, and by visualizations, or more precisely, scholarly figures that act as the lingua franca in this joint enterprise. Figures, like data and processing power, accelerate research as they allow for communication between communities of practice, that internally rely on mutually opaque terminologies, differential equations, algorithms, or distinct workflows (cf. Figure 1).

Like the featured article by Lev Manovich in the last issue, my arguments will be heavily informed by my own expertise, which includes a scholarly background in art history, classical archaeology, psychology, what is now called graph data, and complex network science. My own work addresses questions and challenges of art, architectural, and cultural history, using a multidisciplinary approach that integrates qualitative inquiry and observation, with methods of computation, natural science, and information design. The resulting research processes are mostly characterized by international collaboration and co-authorship. Work procedures are expressed in a distributed, lab-style environment inspired by architectural think tanks, corporate design studios, and labs in physics or systems biology. Products aim at hi-impact journals, conference proceedings, and occasional monographic publications, all of which ideally also cater to a broad audience. Striving to deal with images and figures in the manner of high-quality artist publications, some results of our work are also increasingly themselves exhibited as artworks, which is not by accident. Even though inspired by science, my approach is not without precedence in art and architectural history, standing on the shoulders of practitioners such as Geymüller, Barr, Malraux, Kubler, the Eames, Venturi & Scott-Brown, Doxiadis, Koolhaas, and others.[2]

While I believe that the approach outlined here will become a prevalent model in the ecology of methods aiming to understand art and culture, it is important to mention that what is currently called digital art history is both less and more. Just like systems biology has established itself besides more traditional forms of practice, such as the observation of individual birds, the quantitative multidisciplinary approach will exist beside other forms of art historical practice. Digital art history is less than what I outline below, as many large-scale technological projects in the field are characterized by engineering approaches that aim to build tools to find patterns and facilitate traditional practice, such as the comparison of individual images. Aiming to understand the large-scale patterns we reveal, I will underscore that such engineering needs to be complemented by science, in the sense of physics, i.e. by formalizing quantitative laws.[3] Digital art history is also more than what I outline below, as it includes a large variety of methods that do not require scientific, computational, or aesthetic skills, while being immediately accessible even to non-tech-savvy or science-minded art historians. These immediate aspects include high-bandwidth browsing of source texts, images, and urban environments, ever closer or distant readings, and of course the simple usage of apps, digital libraries, databases, and other exploration tools that continue to be developed over time.[4]

Looking into the future, all these immediate aspects of digital art history will probably revert to simply being called art history, like digital astronomy, after serious debate in the 1980s, reverted back to astronomy, as almost no astronomer could imagine anymore to work without digital data or digital methods.[5] The approach outlined below on the other hand may grow into a systematic science of art and culture, with a similar growth trajectory as systematic biology, "Broadly defined".[6] This systematic approach will also shed the denominator digital over time, as methods may include analog, quantum, and other forms of computation, and data may come in other forms than digital zeros and ones. With that in the back of our mind, for now, we can safely locate the outlined approach within the scope of digital art history.

In the following paragraphs I will argue that the process of art history is both transcending and exponential, while the discipline of art history, in principle, has no limits in method. While the term big data is relative or nonsense, I will show that "more is different" and that understanding the resulting "organized complexity" in art and culture requires an integration of natural science and humanistic inquiry, which is not a reason to fear, but a positive development to embrace. I will convince the reader that humanistic inquiry and natural science share the same basic research pipeline, and that norm data is just the clear end of a massive gradient of uncertainty. Concluding, I will point to outstanding examples of art historical research beyond the discipline of art history.

## The process of art history is transcending

It is not new to point out that the process of art history transcends the boundaries of specialized disciplines dedicated to more or less arbitrary subsets (cf. Figure 2). Salvatore Settis, for example, reminds us that the radical divorce between classical archaeology and modern art history is obviously made up, as we all know that we are dealing with a single historical process.[7] Nevertheless, students of art history are constantly exposed to supposedly necessary limitations. Art history pioneer Heinrich Wölfflin famously said that even though "it is hard to answer the man who regards history as an endless flow", "intellectual self-preservation demands that we should classify the infinity of events with reference to a few results",[8] obviously implying his famous pairs of terms, as well as nations, or stylistic periods. In Wölfflin's tradition, a popular introduction to art history in the German language dedicates several sub-chapters to delineate subject areas of art history as necessary specializations to get a job.[9] The most curious products of such definitions are tenure-track positions for aspiring young faculty that are sometimes limited to extremely narrow topics, such as profane buildings of a particular family, covering a couple of decades in a particular Italian city, analyzed with a particular method. On the other hand, brilliant art historians that have "not specialized enough" are often limited to teaching general survey courses in non-tenured adjunct positions or to act as guides in the tourist industry.

As a systematic science of art and culture transcends such limitations, the exploration and summary of the process of art history, as a whole, becomes a research priority that is again as important and justified as the inquiry into local specifics. Widening the scope it aims to advance our understanding of the history of all made things, in the spirit of George Kubler, and following Marcel Duchamp in asking if there can be any works that are not "of art".[10] As Ernst Gombrich pointed out, in obvious allusion to Charles Darwin, "the coral reef of culture was built by short-lived human beings, but it's growth is a fact not a myth".[11] As with coral reefs in the sea, we can and should study individual subsections of art history, while not forgetting that we might get novel insights looking down at the whole structure and dynamics from space. Due to the organized complexity involved, as detailed below, we will be rewarded with breathtaking beauty and radically new insights that cannot be achieved by local inquiry.

## The process of art history is exponential

The process of art history as a whole is intimidating. World population currently grows faster than exponential, which means the so-called Malthusian explosion is indeed exploding itself.[12] As technological innovation extends the carrying capacity of the planet and raises the amount of artifacts that can be produced by a single individual, the dynamics involved are approaching the fidelity of a global nervous system, whose understanding becomes crucial for our future survival.[13] As the subjects of art history, both past and present, grow with the system and determine parts of it, their study can provide a segue towards a better understanding of the system as a whole.

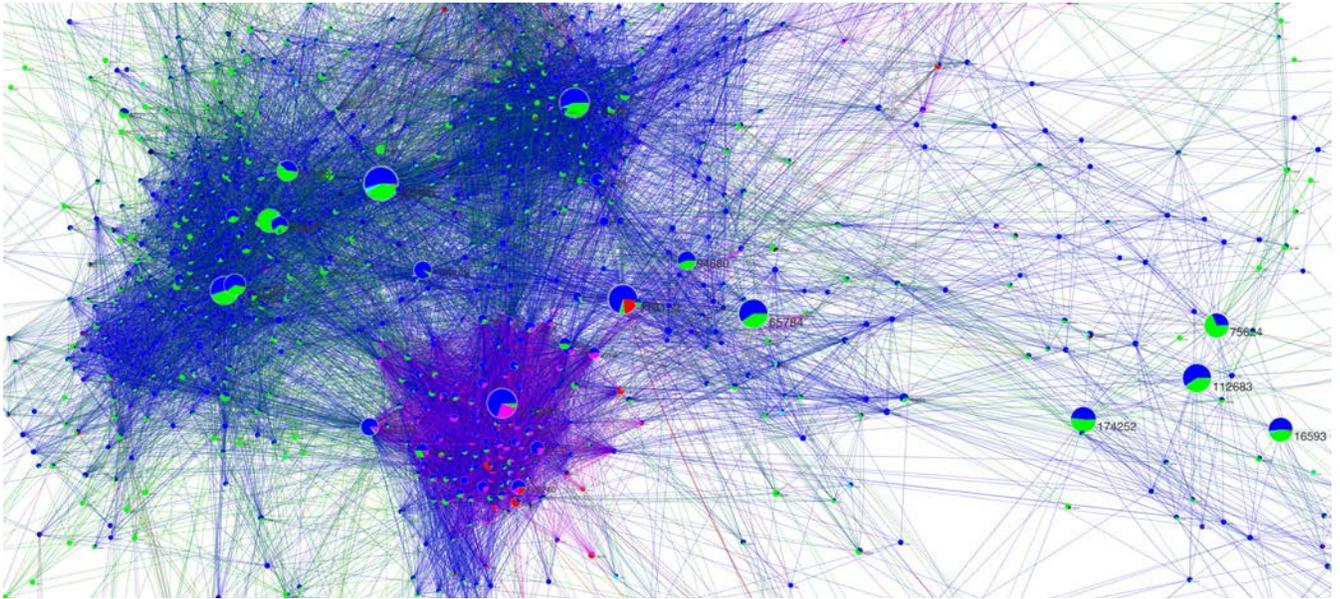

**Figure 2: Subject areas in art and culture are highly entangled.** This becomes visible by using computation and visualization to filter and map emerging communities of topics (nodes) and their overlap (links), across subjects (blue), locations (green), eras (magenta), and individuals (red). The evolution of subject areas and their overlap, as discussed further in Schich and Coscia "MLG 2011", bears striking stabilities over decades, while also indicating non-intuitive growth that needs to be measured, similar to the evolution of the "PACS" classification in physics. Like in classical archaeology (shown here), we can expect similar organized complexity in general art history, with a stronger focus on known individuals, in addition to objects, locations, and eras. Visualization: Maximilian Schich & Michele Coscia.

While the entire history of the Paris salon over more than two centuries comprises less than 130.000 artworks,[14] in 2013 more than 350 million pictures were uploaded by Facebook users every single day.[15] Judging from the fraction of Instagram images identified as self-portraits in the Selfiecity project,[16] this likely means that our daily output eclipses the whole history of portraits, not only as covered by the discipline of Western art history, but from the moment our species started to produce images more than 40.000 years ago[17] to at least well into the 20th century.

The number of known artists, as noted in Allgemeines Künstlerlexikon and the Getty Union List of Artist Names, grows exponentially since about 800 years, on a trajectory that is faster (still) than world population.[18] Today, according to a recent publication by the Inter-American Development Bank, the creative industry or, as they call it, Orange Economy,[19] if it were a country, would be $4.29 trillion dollars in size, i.e. larger than the German economy, only surpassed by Japan, China, and the United States. With $646 billion dollars it would be the ninth-largest exporter of goods, and have the fourth largest labor force with 144 million workers. This means the current labor force in the creative industries eclipses the documented creatives and artists in AKL by about two orders of magnitude. There are over 100 times more creatives making a living today than noted in history since 1200 CE.

Looking at the inventories of well-funded museums, similar growth trajectories would become evident, which either means the number of objects grows more or less exponentially over time, or we tend to forget stuff in an exponential way. No matter how much the actual growth of production, the documentation bias, and the decay of preservation contribute to this situation, there can be no doubt that the exponential nature of our record presents a serious challenge working towards a better understanding of the process of art history.

There is no established way of qualitative inquiry that can deal with this form of dynamics. And there is no trivial way to dissect the exponential growth trajectories into meaningful, non-overlapping periods. As the exponential growth of cultural output, like world population, contributes to a large number of indicators that characterize the sustainability of our species, understanding the process of art and cultural history advances from a harmless hobby horse to a mission critical application of research and education, feeding into a deep history that ties all disciplines into a single narrative of phase transitions from the big bang to our own daily experience.[20]

## The discipline of art history has no limits in method

As we strive to understand the process of art history, we are using established methods and develop new approaches, both qualitative and quantitative, communicating our results to emerging communities of interested scholars, and a broader audience. Ironically, much time is spent to define the in-crowd, to rewrite the creation myth of our practice, to debate on a purely theoretical level, or to reframe the field from individual perspectives. All this is necessary, and this journal consciously provides a forum for such discussion, but we should not forget that our mission is first and foremost to understand the process of art history. Isn't it ironic that a cited search for Warburg's Bilderatlas[21] returns a wealth of literature theorizing the approach, while the majority of practitioners that deals with large amounts of images have never heard about Warburg, even though his idea of Mnemosyne may be as important as the ideas of Planck are for quantum mechanics? As Vitruvius recommends for good architects, we must combine theory and practice to avoid hunting shadows while reaching authority and getting to the substance.[22] Like the architect's goal is to build, our own goal is to understand the historical process. How we call the procedure of reaching this goal is secondary.

Similar to the menu in a Vietnamese restaurant, we are currently confronted with a large variety of concepts, many of which share similar ingredients, while only the initiated are familiar with the subtle and sometimes radical differences. Digital Art History, Digital Humanities, Humanities Computing, Computational Art History, Culturomics, Cultural Analytics, and Data Science in Art History are only some of the pertinent concepts on offer in the naming game that leads up to a major tipping point or phase shift in the system.[24] To achieve relevance towards our aim of understanding the process of art history, and make impact with the audience, it makes no sense to build walls and hide behind one name or the other. It also makes no sense for self-identified "traditional" art historians to avoid, exclude, be afraid of, or look down on those engaged in new perspectives and approaches. Scientists of art and culture will not take over traditional art history. They are not computer-people that provide researchers with a visualization or an automatic tool. They are not a service. Scientists of art and culture are researchers sharing the same goal, namely to understand the subjects and processes of art history. The only difference is that they do not stick with Pad Thai, but go for the entire menu to reach the goal, and if necessary they change restaurant, and learn to cook or even invent their own cuisine.[25]

Since I did my PhD in art history, with "too much" classical archaeology, and joining a physics lab as a post-doc, I have been asked very frequently to define what I consider myself. My usual answer is an anecdote of artist Anish Kapoor being asked if he considers himself British, Indian, or Jewish. His smart reply is to point out that we have to stop compartmentalizing people. Instead of providing one of the concepts above, I point out that my aim is to understand the nature of culture by integrating art history with computation, physics, and information design. I am a Professor in Arts & Technology and a founding member of the Edith O'Donnell Institute of Art History. As such, I am teaching courses in art history as well as courses engaging in cultural data science and information design. My research combines both strains and refuses to limit itself to a particular discipline.

## Big data is relative or nonsense, but more is different

Both my own work and the work of Lev Manovich has been described as dealing with big data, which reflects the size difference of our projects in comparison to other work in art history. We have to admit, however, that we are not overwhelmed by data in the same way as data scientists that deal with real-time streams that are gigabytes per second in size. We do not have to remove or cloak potentially useful data as it comes in. And most of what we do even runs on a single machine, such as the one on your desk. Our data is large, but it would be much larger in an ideal world. In a system of 120.000 individuals moving from birth to death, we have a mere couple of thousand data points over two thousand years, even for the largest centers, such as Paris; in Selfiecity out of 120.000 Instagram images, only 3600 selfies are above the threshold of quality to make it into the visualization.[26] As a consequence, like most quantitative scientists when working towards publication, we are worried about issues of under-sampling and bias, in short about having enough data, not about being overwhelmed by too much. Even with the current explosion of data availability, these issues will remain, and, like in economics, social science, and biology, the discussion of bias will occupy a significant amount of time and effort. The extensive discussion of bias in the Online Supporting Material of our recent paper in Science Magazine is a striking example.[27] On the other hand, the discussion of bias is not a weakness, but a strength of quantification. It is easy of course to observe that minority artists are under-documented, while it is an actionable insight for future funding and research to say by how much compared to the population as a whole.

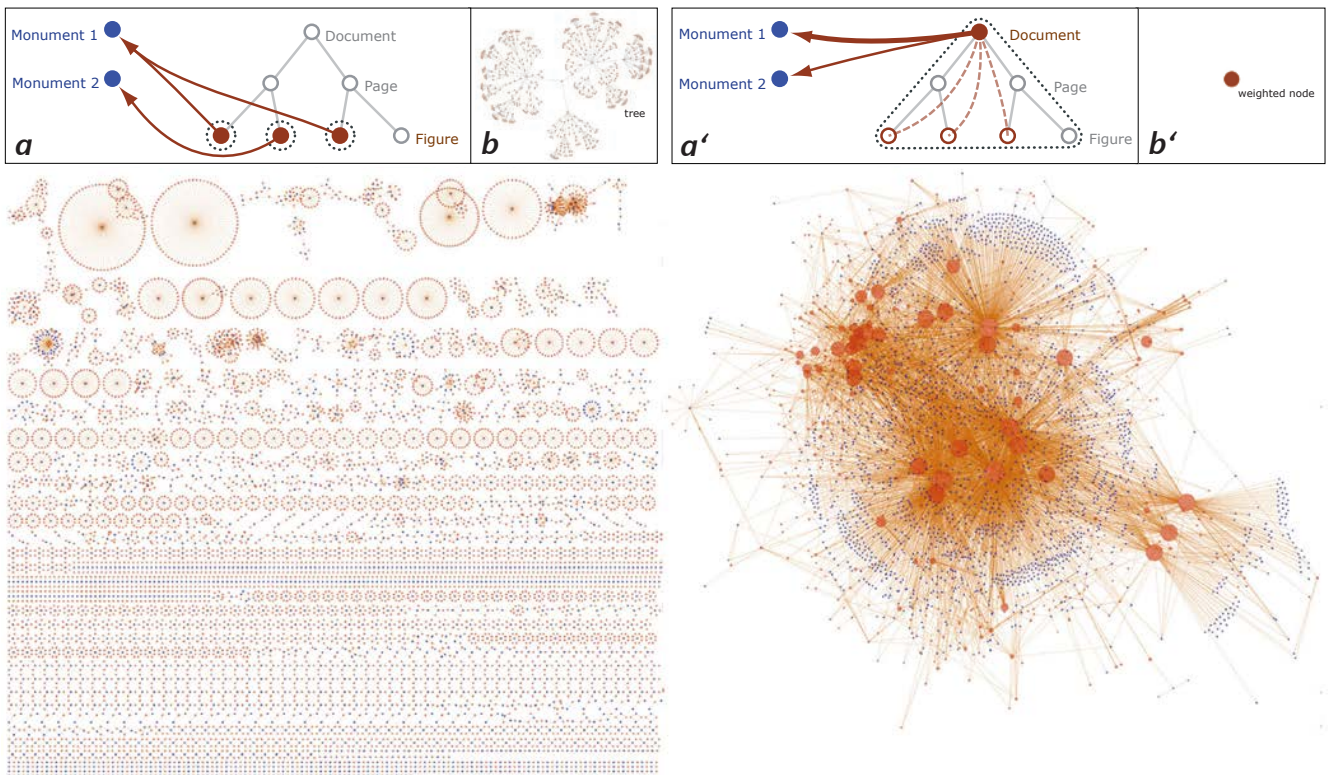

**Figure 3: Organized complexity emerges from aggregates of local specifics.** To the left, modern documents (brown) are connected with ancient monuments (blue) in the "Winckelmann Corpus". To the right, nodes summarize whole documents, integrating individual drawings and text occurrences into books, etc. As a consequence, the system undergoes a so-called phase transition, forming a single connected cluster. Similar to other complex networks, such behavior is the subject of mathematical graph theory and physics. Its observation also has immediate consequences for further funding and research. With the largest cluster spanning 100%, as opposed to an expected 90% (cf. Schich "Revealing Matrices"), the Corpus obviously contains monuments "known by Winckelmann", excluding those "known by his time but not himself". Data: Kunze & Betthausen "Corpus", Visualization: Maximilian Schich.

At its best, the term Big Data is not an absolute, but relative term that should be avoided in practice, even though it may (still) help when journalists use it. Big data is similar to colossal order in architecture. Standing in front of Palazzo Capitanio in Vicenza, the columns indeed seem colossal and intimidating relative to the facade as a whole, while in fact the building is not exactly the size of New St. Peter's in Rome. In a similar way relatively small amounts of data may look intimidating in relation to qualitative methods of inquiry. For an art historian doing the catalogue raisonné of even a very prolific artist, 1 million AKL artists or 1 million Manga pages may seem big.[28] But for data scientists big is when considerable infrastructure is needed to store data, such as 10.000 Tweets per second as they come in, or when they run into the necessity of throwing away data unseen, as in case of the Large Hadron Collider, where too much image data is generated to even store, let alone to fully analyze, while using the best technology available. For those curious, the CMS detector of the LHC produces 40 million images at 1 gigapixel resolution per second, which is more than 25 times the number of images in Prometheus Bildarchiv at the largest resolution available in some select cases within Google Art Project.[29] From that perspective the available amount of digital data in art history is almost ridiculously small.

At its worst, the term Big Data is nonsense. Looking for great literature on the topic, it is useful to compare a Google Books search for "big data", which returns a broad audience book as the top result, while a search for "large data", returns the practical textbook on data science also recommended by Lev Manovich in the last issue of this journal.[30]

Be that as it may, on a practical level, large and eventually really big data is relevant to understand the process of art history as a whole because "more is different".[31] As we cannot imagine the full structure and dynamics of the great barrier reef by looking at a couple of fish or a bunch of polyps, we cannot understand the large-scale structure and dynamics of the process of art history by studying a selection of paintings, artists, or archival records. Like a coral reef, the process of art history is a product of "local activity"[32] by a large number and variety of actors, forming a highly entangled complex system that is literally more than the sum of its parts (cf. Figure 3).[33] The coral reef of culture, like biology, includes large networks of complex networks[34] whose structure and dynamics we can only understand given large amounts of data. The networks involved contain emerging information that is not a property of individual actors, objects, locations, periods, or events, but a property of hard to define aggregates or of the system as a whole. As a consequence, to advance our understanding, we have to combine our traditional domain expertise in art history with methods of complexity science, such as matrix algebra, and advanced graph theory.[35]

## Understanding complexity needs science as well as humanities

Nurturing natural science methods to understand the process of art history promises to overcome the long-standing separation of "nomothetic" law disciplines, such as physics, and "ideographic" event disciplines, such as history, as postulated by Wilhelm Windelband in 1894 and famously lamented by C.P. Snow in the 1950s.[36] Warren Weaver in 1948 and implicitly Jane Jacobs in 1961 have already argued that such an integration is possible and indeed necessary to address abundant problems of "organized complexity", in both economic and urban systems.[37]

In A Network Framework of Cultural History,[38] we implicitly provide a rigorous mutual justification for such an integration of quantitative and qualitative research. The article shows that quantification in the humanities does indeed work by bringing evidence for the physical "laws of migration" spanning over 800 years, based on simple birth and death records of large numbers of artists and other individuals. On the other hand, the article also shows that quantification cannot replace qualitative inquiry, as the system of cultural history is characterized by massive fluctuations on a local level (cf. Figure 4). Both methods of inquiry bring essential ingredients to the table. Delineating examples, the article further promotes the integration of qualification and quantification by revealing sense-making cultural meta-narratives as they emerge from large amounts of granular information, and by helping to cross-fertilize qualitative domain expertise within the context of a big picture. Finally, as mentioned above, the rigorous quantification of bias on mesoscopic and global levels in the Supporting Online Material adds to the general usefulness of the combined approach.

While our Science paper was a major breakthrough, the proposal of integrating humanistic inquiry, computation, natural science, and information design to understand the process of art and cultural history, still often invokes a manifest disbelief in quantification and sometimes the almost insulting conviction that such a proposal can't be much more than "data management". Such reactions are not surprising, as the process of understanding art and culture is still dominated by qualitative humanistic inquiry, and the necessary foundations are not taught within the standard curriculum. Technology within the discipline of art history, including quantitative science, is mostly perceived and treated as a complement or service, where qualitative researchers call in computer experts and designers to support their qualitative inquiry. An example of this phenomenon is the social network diagram published recently at MoMA, also cited by Lev Manovich in the last issue. Intended to improve over the famous original Barr chart, the new MoMA diagram has been marketed as the result of a high profile collaboration between a curator at the museum and network analysts in the business school of Columbia University. While the new diagram could have been done by a reasonably talented undergraduate in Digital Humanities within a few minutes, the original chart's irony, using Picasso's bull as a layout algorithm, could only have been produced by an art historian, such as Alfred Barr, who mastered the production of images just as much as their curation.[39]

While some technological applications in art history have achieved flagship status within university departments and research institutes, starting in the 1980s, the differences of perceived authority are still expressed in salary differences between professors and institute leadership recruited from those doing qualitative inquiry, versus lower-paid adjunct or well-paid but temporary employed computer experts and designers on the other side, notwithstanding their pertinent expertise, often underscored by a PhD in art history. It is a step forward to underline that "humanists must work side-by-side with technical experts [...] to get tools, portals, access, etc.", as Thomas Gaethgens, head of the Getty Research Institute, recently quoted Johanna Drucker.[40] Such acknowledgment breaks with the implicit pattern of subordination but is not enough. Drucker's statement, and indeed the whole definition of digital humanities according to leading practitioners[41] still implicitly assume that the application of technology in art history is an engineering problem, with the final goal to produce means that help the actual researchers doing their inquiry. To achieve a deeper understanding of the process of art history we cannot employ such a procedure, akin to civil engineering, where engineers build the street while working side-by-side with future drivers. Instead, deeper understanding is like the honey in a bee's nest. It can only be reached by those who are able to master and adapt the twig or whatever tool will take them there.[42] Everybody involved in the process must have enough expertise in both arts and technology to at least collaborate towards achieving the ultimate goal of a deeper understanding. It serves to immediately point out that such a proposal is not the suggestion of "white male science" to take over the arts and humanities.[43] Indeed, being modeled on established practices in multidisciplinary network science and systems biology, the proposed science of art and culture promises to attract enthusiasm from a large diversity of researchers, coming from all continents and much better gender balance than discrete communities of practice.[44]

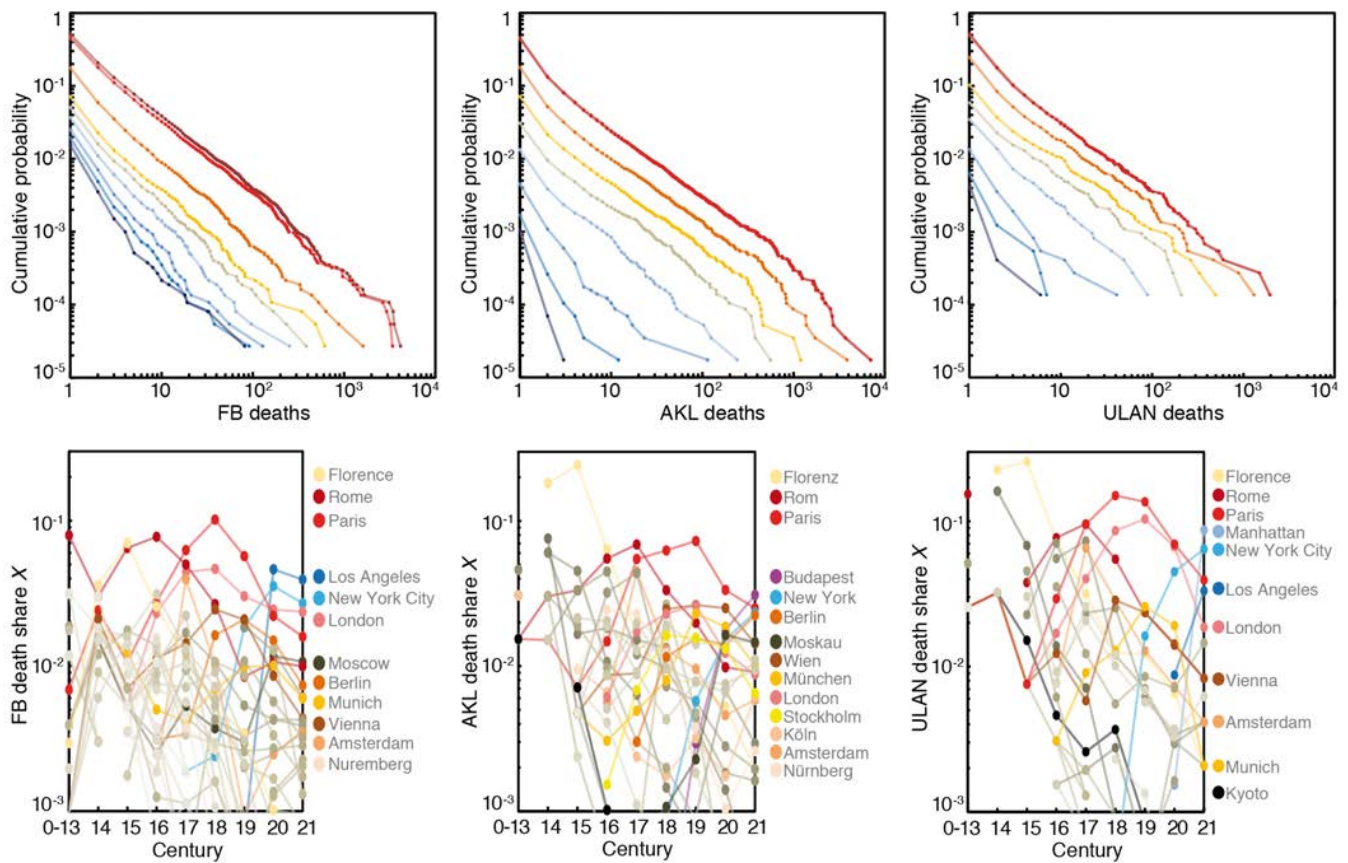

**Figure 4: Quantitative science and qualitative inquiry are both necessary and complement each other.** The three plots above highlight the necessity to quantify physical laws in cultural history, indicating a heterogeneous size distribution of cultural centers that grows more or less exponentially over time while being stable in slope throughout history. The three plots below make a case for qualitative inquiry, by exposing massive fluctuations in the relative share of notable deaths in cultural centers. Both phenomena are consistent across datasets, even though Freebase.com (FB) has very little overlap with Allgemeines Künstlerlexikon (AKL) and the Getty Union List of Artist Names (ULAN). All plots see Schich et al., "A Network Framework of Cultural History," including the Supporting Online Material.

Uri Alon, author of a popular Introduction to Systems Biology, has introduced a striking model called the "cloud of uncertainty", which can help us to clarify the difference between engineering problems and problems of science.[45] Projects that aim to build tools, portals, and access are engineering problems as they aim to go from problem A to an imagined future solution B. Examples include the digitization of all books ever published, or a database of all paintings in public collections. Both applications require highly skilled researchers, masterful coordination, sophisticated technology, and efficient workflows to be successful. The results may be highly useful to traditional practitioners, but in themselves do not necessarily contribute to a deeper understanding of the subject matter. Projects that aim towards such a deeper understanding on the other hand, may include some engineering, but are very different in nature, no matter if they choose to employ qualitative or quantitative methods. They need to go where nobody has gone before, even in imagination. The difference is that starting with a situation A, we may find out that the imagined solution B is unachievable, putting us into the "cloud of uncertainty", from which we can only escape by mastering whatever method is necessary to reach an unknown and maybe surprising solution C. In addition to scientific skills, this may involve to overcome negative emotions and depression towards reaching the happiness of insight.[46] As Paul Feyerabend pointed out, this enterprise is essentially anarchic, and we have to act like undercover agents, who play the game of reason, to undercut the authority of reason.[47] There is no fixed workflow pipeline or service that we can call in like a construction firm in civil engineering. Instead we are required to learn, master and adapt our methods and tools as we go along.

Aiming towards the unknown to eventually find surprising insight is a common trait of basic science and research (Grundlagenforschung), no matter if qualitative or quantitative. Of course, while there is no fixed workflow in this enterprise, there are general recipes and procedures that help to formalize the process of inquiry and raise the chances for new insights. Hedging our resources like an angel investor or venture capitalist, with collaborators being involved in multiple projects, we can minimize the risk and ensure the overall success of a given group of researchers.

## Humanistic inquiry and science share the same basic pipeline

A systematic science of art and culture will align with traditional qualitative scholarship in art history, not only in terms of question, but also in terms of workflow, fixing a major shortcoming in established digital practice within the arts and humanities. Over decades we have spent a large amount of energy to develop data models and standards based on formal logic and anecdotal evidence.[48] This was important to get digitization and digital workflows off the ground, but violates a basic principle of scholarship, as it is impossible to arrange material without prior collection and observation of its actual structure.

Preparing an individual piece of scholarship, such as a catalogue entry, a journal article, or a book-sized monograph, we would more or less intuitively follow Cicero's sequence of inventing a speech: First we would collect material, then we would arrange the material, formalize the story, if there is one, and finally deliver it to our audience.[49] In decades of large-scale database projects we have essentially violated this sequence by arranging the material based on our expectations or sometimes ideology, as opposed to taking a deep look into all the material once it is collected, either by using our own eyes or if necessary more sophisticated measurement instruments. Presupposing large-scale structure to be average, random, or intuitive, many database projects were content to create search platforms or browsing tools, whose aim was to facilitate traditional qualitative inquiry. As the emerging organized complexity in the collected material often was out of sync with presupposed expectations and intuitions, it is not a surprise that many large database projects failed to attract a wider and more persistent audience of users.

Quantitative inquiry that aims to map, understand, and explain organized complexity in large collections of data provides a remedy in this situation. Like in the human genome project, where the successful collection of data did not bring an immediate cure for cancer but did start a whole new field of inquiry, decades of digital data collection in art and culture provide a highly promising point of departure. In fact, due to systematic quantitative inquiry, the decade long effort will finally stand up to its promise. Once we know and understand the emerging complexity, we will be able to pose and address new qualitative and quantitative questions, which we can't even imagine today. Like in other areas of data-driven science, the resulting pipeline will resemble Cicero's basic sequence of invention, enriched by infinite feedback, as formalized by leading data scientists and designers.[50] Indeed, one could imagine the resulting pipeline with feedback as an auto-catalytic cycle of research breeding more research, with quantification as an accelerating enzyme (cf. Figure 5).

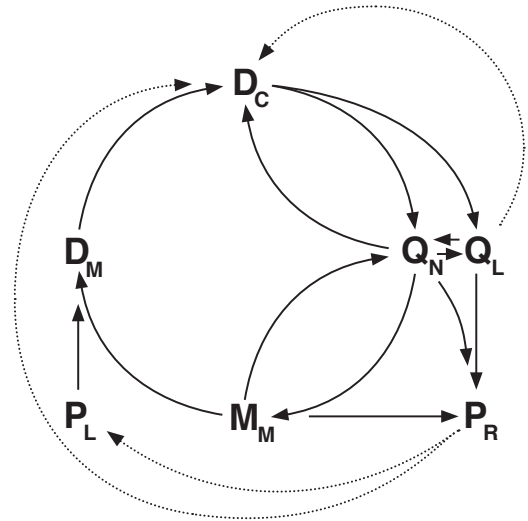

**Figure 5: Quantitative science catalyzes the established sequence of digital scholarship.** Initially, data models ($D_M$) are mostly defined using philosophy ($P_L$), in particular formal logic based on anecdotal evidence. Traditionally, this leads to efficient data collections ($D_C$), new qualitative observations ($Q_L$), and eventually the publication of results ($P_R$), which in turn may lead to better data collection, but usually leaves the original data model intact. Quantitative measurement ($Q_N$) of organized complexity closes the loop as it leads to creation of mathematical models ($M_M$), which lead to accelerated change of data models, data collections, and more novel insight. Domain expertise, computation, and visualization are necessary throughout the process. Of course, the figure, inspired by the Eigen and Schuster "Hypercycle", is a cartoon crying for measurement itself. Image: Maximilian Schich.

As such, qualitative and quantitative practice will feed into a common cognitive process that will advance our understanding of art and cultural history. As there will be variations in procedure, many papers will start with a figure explaining the pipeline.

## Norm data is just the clear end of a massive gradient of uncertainty

A good example for the need of quantitative measurement is our obsession with norm data, authority files, and data model standards. Almost nothing in art history is normal, in the sense of a normal distribution with a sense-making average, as in case of a Gaussian bell-curve containing more or less average examples around it. There is no average artist, no typical triumphal arch, no regular Roman sculpture, and no normal nativity scene. Wherever we look we usually find one or a handful of exceptional examples, and a more or less long or "fat tail" of irregular or hybrid examples that are not-so-well-documented, not-so-typical-looking, not-so-well-preserved, or not-so-easy-to-attribute.[51]

I am not saying that there are no well defined groups of objects. What I am saying is that, based on existing evidence, we have to deal with massive gradients of uncertainty. If we want to understand the art market, beyond some well-identified paintings, collectors, auction houses, etc., we have to deal with a vast majority of uncertain attributions, a majority of rare and unknown actors, and of course the unknown amount of "dark data". In other words, art history has to deal with probability distributions and potential sources of bias, just like social science, biology, and other quantitative fields. It is an illusion to think an editorial process can combine normed classification and addressability. If you run an archive: Do assign identifiers to your records, optionally do crude classification, and leave "figuring out" to the whole community of researchers.

To give an example: In the last two years the incredibly talented computer scientist John Resig, who gave us jQuery and Processing.js, is essentially touring prominent visual resource collections in art history to apply computer vision algorithms in order to find duplicate photographs of artworks. Called in like a service, the premise is "to change how photographs and images are managed in archives, libraries, and museums",[52] working towards a unified or normalized collection of photos that will facilitate traditional scholarship. Such improvement of management by engineering is important, but loses an important chance to accelerate the science of art and culture. As every art historian knows from their own specific practice, highly similar images that can be matched like copies are just the most simple case of visual family resemblence.[53] So one must ask, if we really should split computational management of visual resource collections from scholarly inquiry in art history. Wouldn't it make much more sense to publish the photo archives, like the human genome, to facilitate an explosion of quantitative research not only into duplicates and near-duplicates, but into the entire gradient of similarity? To clarify the potential: A groundbreaking and highly relevant paper, published in the area of computer vision as recent as 2012, already has more than 2500 citations,[54] which means there are likely hundreds of groups that would be more than happy to work with image data that spans more than the last 20 years.

## There is outstanding Art History beyond Art History

It is easy to cite more such examples that are currently beyond the radar of the discipline of art history. Like social network analysis was beyond the radar of physicists in complex network science 15 years ago,[55] there is a vast amount of work that either precedes or runs parallel to current efforts in digital art history.

In 1967, French geographer Jacques Bertin published his Semiology of Graphics,[56] which, if I had to choose, would be the one book on data visualization that I would take to Mars, if I had to leave everything else behind. Introducing matrix permutation, he claims algorithmic analysis has to go hand in hand with manual sorting. He demonstrates this by using a classification of Merovingian artifacts, i.e. an example taken from the realm of archaeology and art history, likely from a stream of research that discussed the pros and cons of dimensionality reduction, such as Principal Component Analysis, since more than 50 years ago.[57]

Lev Manovich's image plot software is preceded by a contribution in computer science, published in 1996,[58] just like my own frequency distributions of ancient monuments in Renaissance documents are preceded by Heinrich Dilly, who counted the frequency of artists in the titles of art historical literature over several decades, publishing the result almost ironically in a volume on art history and the Frankfurt school of philosophy.[59] Stanley Milgram did word clouds with a sense-making layout 30 years before they took off, and computer linguists are jealous of James Joyce for having implicitly outlined almost any possible question.[60] We should appreciate and cite such colleagues and giants on whose shoulders we stand. But of course we should also be aware that we can go much further than we could ever before, thanks to unprecedented amounts and quality of data, as well as advances in computational power and scientific method.

Only since recently it is possible to get and deal with millions of tourist photos plus imagery taken from Google Street view, to extend theories of perception aesthetics by mapping the density of tourist attention and even calculate the density of viewing cones of individual tourists, as a side effect of reconstructing buildings in 3D without human interaction.[61] Only since recently we can use algorithms to convincingly date architectural details, in order to map the evolution of palace facades in Paris, strikingly mimicking the perception of a well-trained art historian strolling through the city.[62] Only with services the size of Facebook, it has become possible to study the spreading of visual memes on a large scale, revealing cascades that resemble a mathematical theory of biological evolution.[63] Trend analysis in fashion, which traditionally bears striking resemblance with scholarship in art history, is increasingly driven by larger sets of data and quantification.[64] Finally there is an increasing amount of analysis into paintings and artworks, done and published by natural scientists in multidisciplinary environments.[65]

# Conclusion

In this article I outlined a perspective for a systematic science of art and culture that integrates qualitative inquiry with computation, natural science, and information design. As such, cultural science shares the aim of understanding the process of art history with so-called traditional practice. It explores unknown complex emerging structure and dynamics by analyzing large data sets, using both quantitative measurement and qualitative inquiry. Similar to systems biology, the procedure is characterized by multidisciplinary co-authorship and publications that make extensive use of scholarly figures.

The Journal of Digital Art History has the potential to fill an important gap in this enterprise. Positioning itself in a disciplinary niche within an emerging journal hierarchy,[66] similar to Nature Physics, the Journal of Digital Art History complements existing journals that mediate between art and science, such as Leonardo, and multidisciplinary journals, such as Palgrave Communications, the new social science and humanities equivalent of Nature Communications. The emergence of such a publication infrastructure provides important opportunities for students and researchers engaging in an art and cultural history without limits. With an estimated market demand of more than 140.000 data scientists,[67] and a growing abundance of cultural data, there can be no doubt that the laboratories engaged in the science of art and culture will have an important function in society and are bound to thrive.

# Notes

[1] Acknowledgments: The author wishes to thank the anonymous donors of UT Dallas ATEC Fellowship #1 and Dirk Helbing at the Professorship for Computational Social Science at ETH Zurich for their generosity and hospitality. As this article is a perspective, not a map, it fails to mention all the excellent work to be summarized in a hopefully forthcoming review.

[2] Regarding pioneers see Cortjaens and Heck, Stil-Linien Diagrammatischer Kunstgeschichte on Geymüller; Schmidt-Burkhardt, Stammbäume der Kunst on Barr; and Schich, Rezeption und Tradierung als komplexes Netzwerk pp. 156-160 on Kubler; In addition, compare the workflows leading to Malraux, Le Musée imaginaire de la sculpture mondiale; Eames and Eames, Powers of Ten; Venturi, Scott Brown, and Izenour, Learning from Las Vegas; Koolhaas and Office for Metropolitan Architecture, Content; See also Wigley, "Network Fever," on Doxiadis etc.

[3] Compare Richard Feynman's definition of "understanding" in Feynman, Six Easy Pieces pp. 24/25.

[4] High-bandwidth browsing cf. "Google Images;" "Google Earth;" and "Google Books." Ever closer and distant readings (Moretti and Piazza, Graphs, Maps, Trees.) see for e.g. gigapixel images in "Art Project - Google Cultural Institute;" versus Crandall et al., "Mapping the World's Photos."

[5] Roger Malina, "Yes Again to the End of the Digital Humanities ! Please !"

[6] "Our Approach | Broad Institute of MIT and Harvard."

[7] Settis, "L'opera Di Paul Zanker E Il Futuro Dell'archaeologia Classica."

[8] Wölfflin, Principles Of Art History.

[9] Belting et al., Kunstgeschichte.

[10] Kubler, The Shape of Time; Duchamp et al., À L' Infinitif = In the Infinitive.

[11] Gombrich, The Sense of Order, p. 209; Bredekamp, Darwins Korallen.

[12] Johansen and Sornette, "Finite-Time Singularity in the Dynamics of the World Population, Economic and Financial Indices."

[13] Helbing, "Globally Networked Risks and How to Respond."

[14] Whiteley, Index to the Paris Salon Catalogues.

[15] "Facebook Has a Quarter of a Trillion User Photos."

[16] Manovich et al., "Selfiecity."

[17] Aubert et al., "Pleistocene Cave Art from Sulawesi, Indonesia."

[18] Schich et al., "A Network Framework of Cultural History." Figs. 1A.

[19] Restrepo and Márquez, "The Orange Economy."

al., "Mapping the World's Photos."

[62]Lee et al., "Linking Past to Present."

[63]Cheng et al., "Can Cascades Be Predicted?"

[64]"Fashion Trends for Spring 2015 as Told by Google Data."

[65]Kim, Son, and Jeong, "Large-Scale Quantitative Analysis of Painting Arts;" compares well to Rosa and Suárez, "A Quantitative Approach to Beauty. Perceived Attractiveness of Human Faces in World Painting."

[66]Palla et al., "Hierarchical Networks of Scientific Journals."

[67]Manyika et al., "Big Data."

Maximilian Schich is an Associate Professor in Arts & Technology and a founding member of EODIAH, the Edith O'Donnell Institute of Art History at the University of Texas at Dallas. In summer 2015, he also was a Visiting Scientist at ETH Zurich in Dirk Helbing's Computational Social Science group, where he wrote parts of this article.

He is the first author of A Network Framework of Cultural History (Science Magazine, 2014) and a lead co-author of the animation Charting Culture (Nature video, 2014). He has visualized networks of complex networks in art research (O'Reilly 2010), and analyzed antique reception and visual citation as complex networks (Biering & Brinkmann, 2009). He is an Editorial Advisor for Arts, Humanities, and Complex Networks at Leonardo Journal, and an Editorial Board member at DAH-Journal and Palgrave Communications, the new humanities and social science equivalent of Nature Communications. He has been invited to SciFoo, DLD*, Edge.org, and the Lincoln Center Global Exchange. His most recent work received global press coverage in 28 languages.

Correspondence e-mail: maximilian@schich.info